%% file: PIMRC_conf.tex
\documentclass[letterpaper,conference]{IEEEtran}
\usepackage[letterpaper, left=1in, right=1in, bottom=1in, top=0.75in]{geometry}

\usepackage[utf8]{inputenc}
\usepackage{cite}
\usepackage{amsmath,amssymb,esint}
\usepackage{graphicx,xcolor}
\usepackage[super]{nth}
\usepackage{psfrag}
\usepackage{dotlessi}
\usepackage{pbox,ragged2e}
\usepackage{balance}
\usepackage{placeins}

\usepackage{graphicx}
\usepackage[caption=false,font=footnotesize]{subfig}
\usepackage{gensymb}

\newif\ifstuff
\stufffalse

\input{Macros}

\IEEEoverridecommandlockouts\IEEEpubid{\makebox[\columnwidth]{ 978-1-5386-3531-5/17/\$31.00~\copyright~2017 IEEE \hfill} \hspace{\columnsep}\makebox[\columnwidth]{ }}
\begin{document}

\title{Robust Near-Field 3D Localization of an Unaligned Single-Coil Agent Using Unobtrusive Anchors}
\author{%
\IEEEauthorblockN{Gregor Dumphart, Eric Slottke, and Armin Wittneben} 
\IEEEauthorblockA{Communication Technology Laboratory, ETH Zurich, 8092 Zurich, Switzerland\\
Email: \{dumphart,  slottke, wittneben\}@nari.ee.ethz.ch}}
\maketitle

\begin{abstract}
\input{sections/00-Abstract}
\end{abstract}

\section{Introduction}
\label{sec:intro}
\input{sections/01-Intro}

\section{System Model}
\label{sec:model}
\input{sections/02-Model}

\section{Cram\'{e}r-Rao Lower Bound}
\label{sec:crlb}
\input{sections/03-CRLB}

\section{CRLB-Based Performance Evaluation}
\label{sec:eval}
\input{sections/04-Eval}

\section{Estimation Algorithms}
\label{sec:algos}
\input{sections/05-Algos}

\section{Summary \& Outlook}
\input{sections/99-Concl}

%\clearpage
\appendices
%\numberwithin{equation}{section}%
\section{}
%\section{Link Circuit Model}
\label{app:circuit}
\input{sections/A1-Circuit}

\section{}
%\section{Geometric Signal Gradients}
\label{app:gradient}
\input{sections/A2-Gradient}

\section*{Acknowledgment}
%The authors would like to thank Wolfgang Utschick of Technische Universit\"at M\"unchen for suggesting alternating estimation of position and orientation at IHS 2016, which inspired the proposed algorithm.
We would like to thank E. Riegler and M. Tschannen for valuable comments as well as W. Utschick of Technische Universit\"at M\"unchen for suggesting alternating estimation of position and orientation at IHS 2016.

%\FloatBarrier
%\balance

\IEEEtriggeratref{0}
\bibliographystyle{IEEEtran}
\bibliography{IEEEabrv,BibGregor}

\end{document}

%% file: Macros.tex
\newcommand{\f}[2]{\frac{#1}{#2}}
\newcommand{\fp}[2]{\f{\partial #1}{\partial #2}}

\newcommand{\Tr}{^\text{T}}

\DeclareMathOperator*{\argmax}{arg\,max}
\DeclareMathOperator*{\argmin}{arg\,min}
\DeclareMathOperator{\diag}{diag}

  \newcommand{\p}{\mathbf{p}}
\renewcommand{\o}{\mathbf{o}}
\newcommand{\ag}{_\mathrm{ag}}
\newcommand{\anc}{_\mathrm{anc}}
\newcommand{\pag}{\p\ag}
\newcommand{\oag}{\o\ag}
\newcommand{\pan}{\mathbf{p}_n}
\newcommand{\oan}{\mathbf{o}_n}

\newcommand{\coeff}{\rho}
\newcommand{\s}{s_n}
\newcommand{\y}{y_n}
\newcommand{\w}{w_n}

\renewcommand{\d}{d_n}

\newcommand{\eNoIdx}{\mathbf{e}}
\newcommand{\e}{\eNoIdx_n}

\newcommand{\banNoIdx}{\boldsymbol{\beta}}
\newcommand{\ban}{\banNoIdx_n}
\newcommand{\D}{\mathbf{D}}
\newcommand{\B}{\mathbf{B}}
\newcommand{\eye}{\mathbf{I}}

\DeclareMathSymbol{\jj}{\mathalpha}{letters}{"11}

\newcommand{\wvec}{\mathbf{w}}

\newcommand{\jh}{{\mathbf{j}}_\p}
\newcommand{\jpag}{{\mathbf{j}}_{\pag}}

\newcommand{\param}{\psi}
\newcommand{\bp}{{\boldsymbol\param}}
\newcommand{\FIM}{\pmb{\mathcal{I}}}

\newcommand{\AlgoFull}{{Weighted Least Squares}}
\newcommand{\AlgoLower}{{weighted least squares}}
\newcommand{\AlgoShort}{WLS}

%% file: sections/00-Abstract.tex
The magnetic near-field provides a suitable means for indoor localization, due to its insensitivity to the environment and strong spatial gradients. We consider indoor localization setups consisting of flat coils, allowing for convenient integration of the agent coil into a mobile device (e.g., a smart phone or wristband) and flush mounting of the anchor coils to walls. 
In order to study such setups systematically, we first express the Cram\'{e}r-Rao lower bound (CRLB) on the position error for unknown orientation and evaluate its distribution within a square room of variable size, using 15$\,\times\,$10\,cm anchor coils and a commercial NFC antenna at the agent. Thereby, we find \mbox{cm-accuracy} being achievable in a room of 10$\,\times\,$10$\,\times\,$3 meters with 12 flat wall-mounted anchors and with 10$\,$mW used for the generation of magnetic fields. 
Practically achieving such estimation performance is, however, difficult because of the non-convex 5D likelihood function. To that end, we propose a fast and accurate \AlgoLower{} (\AlgoShort{}) algorithm which is insensitive to initialization. This is enabled by effectively eliminating the orientation nuisance parameter in a rigorous fashion and scaling the individual anchor observations, leading to a smoothed 3D cost function.
Using \AlgoShort{} estimates to initialize a maximum-likelihood (ML) solver yields accuracy near the theoretical limit in up to 98\% of cases, thus enabling robust indoor localization with unobtrusive infrastructure, with a computational efficiency suitable for real-time processing.

%% file: sections/01-Intro.tex
Accurate indoor localization is a highly desired application in biomedical and industrial sectors, for assisted living, access control, the Internet of Things, and smart homes. The most popular wireless indoor localization schemes are based either on time-difference-of-arrival (TDOA) or received-signal-strength (RSS) metrics. However, TDOA systems require wideband transceivers and high-complexity schemes for synchronization and resolving non-line-of-sight bias, while RSS-based localization is usually heavily impaired by fading, shadowing, and antenna patterns \cite{GeziciSPM2005}. %  % TDOA \cite{Dardari2009}RSSI \cite{Dong2012}
The magnetic near-field, while associated with notoriously high path loss, however exhibits useful properties for localization: it penetrates most materials without interaction, thus rendering received signals insensitive to the (typically time-variant) indoor environment \cite{MarkhamSJ2012,SheinkerTIM2013}. 
In addition, the magnetic near-field decays quickly with distance, allowing highly accurate ranging at close distances. Therefore, the near-field has been considered as physical layer for localization on the $10\,\mathrm{m}$-scale in harsh propagation environments, e.g., underground \cite{MarkhamSJ2012} and indoor \cite{AbrudanJSAC2015,SheinkerTIM2013,KyprisTGRS2016,PaskuTIE2016}. 
Most existing work considers tri-axial coil arrays \cite{MarkhamSJ2012,AbrudanJSAC2015,SheinkerTIM2013,KyprisTGRS2016} whose form factor and hardware complexity are however undesired for many applications. In contrast, we assume an unobtrusive setup consisting of planar coils, allowing for an integrated printed coil at the agent and anchor coils which can be flush-mounted on walls without obstructing any activities in the room. 

Position estimation from induced voltages in single-coil setups with unknown agent orientation has previously been studied at smaller scales (e.g., wireless localization of endoscopic capsules) whereby standard solvers for cost function minimization have been applied \cite{Schlageter2001,Hu2005,SongHu2009}. These schemes face degradation due to non-convexity \cite{Hu2005}.\\[-3mm]

\textbf{Contribution of this work:}
We derive the Cram{\'e}r-Rao lower bound (CRLB) on the position error for active near-field 3D localization with unknown agent orientation, based on a sensible coupling and noise models. Therewith, we study performance regimes and demonstrate cm-accuracy localization being achievable in a $10\,\mathrm{m} \times 10\,\mathrm{m} \times 3\,\mathrm{m}$ room with an unobtrusive setup of 12 anchors, using one-shot voltage measurements ($2\,\mathrm{ms}$) and $10\,\mathrm{mW}$ for the magnetic field generation.
We present a rigorous method for estimating the agent orientation given a position hypothesis which effectively reduces the problem from 5D to 3D. Furthermore, in order to combat the problem of high dynamic signal range, we introduce a distance-dependent scaling that results in a cost function relaxation. On this basis, we propose a robust and efficient \AlgoLower{} (\AlgoShort{}) algorithm and show that the cascade of \AlgoShort{} and a maximum-likelihood (ML) solver robustly performs near the theoretical limit with particularly low computational cost.\\[-3mm]

\textbf{Related work:}
The joint estimation of position and orientation of a dipole-like magnet through distributed sensors, each measuring one field component, was studied in \cite{Schlageter2001,Hu2005,SongHu2009}. % for gastrointestinal medical applications (all but Schlageter)
In \cite{Li2008Jour}, a medical microrobot estimates its position and orientation from voltages induced in its near-field antenna due to eight active anchors. For the 5D non-linear least squares problem associated with these works, the Levenberg-Marquardt (LM) algorithm was identified as a suitable solver \cite{Schlageter2001,Hu2005}. The authors of \cite{Hu2005} emphasized the importance of an accurate initial guess for LM because of local cost function minima. The magnetic field Jacobian was provided to the LM algorithm in \cite{SongHu2009} for performance enhancement. Most papers on near-field localization employ the dipole approximation, e.g., \cite{Schlageter2001,Hu2005,SongHu2009,SheinkerTIM2013,PaskuTIE2016,AbrudanJSAC2015,KyprisTGRS2016}. For coplanar 2D localization of a passive LC resonator, least squares estimation errors were compared to the CRLB in \cite{SlottkeVTC2014}. In distinction from planar coil setups, the use of tri-axial coil arrays at the anchors and/or the agent allows for simpler localization schemes \cite{HuMengTM2007,HuSongTM2012,SheinkerTIM2013,AbrudanJSAC2015}. In particular, \cite{HuSongTM2012} uses a simplified localization algorithm to initialize the LM solver applied to the original non-linear least squares problem.\\[-3mm]

\textbf{Organization of the paper:} Section \ref{sec:model} establishes the employed signal and coupling model, setup geometry, and notation.
The near-field position estimation problem is treated in Section \ref{sec:crlb} in terms of likelihood function and Fisher information, which yields the CRLB. Section \ref{sec:eval} presents a CRLB-based evaluation of indoor positioning accuracy with realistic parameters. In order to achieve the postulated accuracy, we derive a novel algorithm in Section \ref{sec:algos} and demonstrate its great advantages in terms of robustness and convergence speed.

%% file: sections/02-Model.tex
\begin{figure}[!ht]
	\centering
  \psfrag{Ag}{\hspace{0.5mm}\raisebox{-1.9mm}{\footnotesize \textcolor[rgb]{.855,.643,.122}{Agent}}}
  \psfrag{An1}{\hspace{-2.0mm}\raisebox{-2.2mm}{\footnotesize $N=12$}}
  \psfrag{An2}{\hspace{-2.0mm}\raisebox{-1.7mm}{\footnotesize Anchors}}
  \hspace{3mm} 
  \includegraphics[width=.85\columnwidth,trim=330 237 280 208,clip=true]{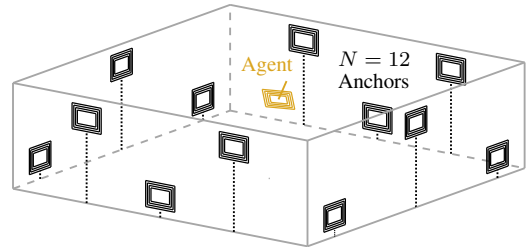}
  \vspace{-2.5mm}
  \caption{Localization setup in a square room, showing the agent coil and \mbox{$N = 12$} anchor coils flush-mounted on the room walls. The coils are not drawn to scale.}
  \label{fig:CoilSetup}
  \vspace{-2mm}
\end{figure}

Throughout the paper, we consider a localization setup in a simple, square room with planar anchor coils installed on the side walls. An exemplary setup is shown in Figure~\ref{fig:CoilSetup} for $N = 12$ anchors. For simplicity we consider anchor coils of equal geometry and parameters. The agent may be located anywhere within the room with arbitrary coil orientation. As agent we consider a battery-driven device which uses reference power $P_\mathrm{t}$ to generate a magnetic AC near-field. The resulting signals observed at the $N$ anchors, which are due to magnetic induction, allow inference about the agent position.

A circuit-theoretic account of these magneto-inductive wireless links, assuming weak coupling and an interference-free environment, is given in Appendix \ref{app:circuit}. It yields the real-valued additive white Gaussian noise (AWGN) signal model
\begin{align}
\y &= \s + \w \ , & n &= 1, \ldots, N
\label{eq:MeasModel}
\end{align}
where the quantities have unit $\sqrt{\mathrm{W}}$. The thermal noise $w_n \overset{\text{i.i.d.}}{\sim} \mathcal{N}(0,\sigma^2)$ has power $\sigma^2 = N_0 B$ with bandwidth $B$ and noise spectral density $N_0 = k_\mathrm{B}T F$ with Boltzmann constant $k_\mathrm{B}$, temperature $T$, and noise figure $F$.

\begin{figure}[!ht]
	\centering
  \psfrag{dn}{\hspace{.4mm}\raisebox{.3mm}{$\d$}}
  \psfrag{en}{\hspace{-0mm}\raisebox{0.4mm}{\textcolor[rgb]{0,.6,0}{$\e$}}}
  %\psfrag{MnJn}{\hspace{-1mm}\raisebox{.5mm}{\textcolor[rgb]{0,.6,0}{$M_n, J_n$}}}
  %\psfrag{MnJn}{\hspace{-0mm}\raisebox{.5mm}{\textcolor[rgb]{0,.6,0}{$J_n$}}}
  \psfrag{bn}{\hspace{-3.1mm}\raisebox{-4.1mm}{\textcolor[rgb]{0,0,1}{$\ban$}}}
  \psfrag{og}{\hspace{.5mm}\raisebox{-2.5mm}{$\oag$}}
  \psfrag{on}{\hspace{-0.9mm}\raisebox{-0.1mm}{$\oan$}}
  \psfrag{pg}{\hspace{-.5mm}\raisebox{-.2mm}{\textcolor[rgb]{1,0,0}{$\pag$}}}
  \psfrag{pn}{\hspace{-0.6mm}\raisebox{.6mm}{\textcolor[rgb]{1,0,0}{$\pan$}}}
  %\psfrag{pn}{\hspace{-.9mm}\raisebox{.2mm}{\textcolor[rgb]{1,0,0}{$\pan$}}}
  \hspace{.15\columnwidth}
  \includegraphics[width=.75\columnwidth,trim=35 10 0 0,clip=false]{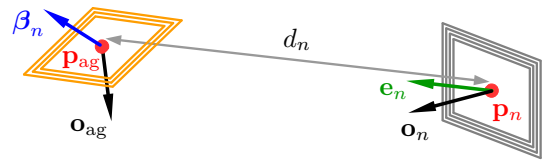}
  \hspace{.15\columnwidth}
  \vspace{-2.5mm}
  %%%%%\\
  %%%%%\subfloat[Abstracted AWGN signal model with $\sigma^2 = N_0 B$.]{%
  %%%%%%\psfrag{r}{\hspace{-2.1mm}\raisebox{0mm}{\textcolor[rgb]{1,.616,0}{$\coeff$}}}
  %%%%%%\psfrag{c}{\hspace{-4.5mm}\raisebox{.7mm}{\textcolor[rgb]{0,.6,0}{$\d^{-3} \J$}}}
  %%%%%%\psfrag{s}{\hspace{1.7mm}\raisebox{1mm}{\textcolor[rgb]{0,0,1}{$\s$}}}
  %%%%%%\psfrag{w}{\hspace{-5.9mm}\raisebox{.7mm}{\textcolor[rgb]{1,0,0}{$\w \sim \mathcal{N}(0,\sigma^2)$}}}
  %%%%%%\psfrag{y}{\hspace{.6mm}\raisebox{0mm}{\textcolor[rgb]{1,0,1}{$\y$}}}
  %%%%%\psfrag{r}{\hspace{-2.1mm}\raisebox{0mm}{$\coeff$}}
  %%%%%\psfrag{c}{\hspace{-4.5mm}\raisebox{.7mm}{$\d^{-3} \J$}}
  %%%%%\psfrag{s}{\hspace{1.7mm}\raisebox{1mm}{$\s$}}
  %%%%%\psfrag{w}{\hspace{-5.9mm}\raisebox{.7mm}{$\w \sim \mathcal{N}(0,\sigma^2)$}}
  %%%%%\psfrag{y}{\hspace{.6mm}\raisebox{0mm}{$\y$}}
  %%%%%\hspace{1cm}
  %%%%%\includegraphics[width=.6\columnwidth]{figures/SigModel}
  %%%%%\label{fig:SignalModel}}
  \caption{Link geometry between the agent coil and the $n$-th anchor coil, explaining all quantities relevant to the dipole coupling model.}
  \label{fig:CoilGeometry}
\end{figure}

A coupling model must be chosen in order to assign geometrical meaning to $\s$. We employ the dipole model, an approximation which is accurate for coupled loops separated by several multiples of the involved coil diameters \cite{DumphartPIMRC2016}. The model relies on the 3D geometric quantities depicted in Figure~\ref{fig:CoilGeometry}: position $\pag$ and orientation $\oag$ of the agent coil, position $\pan$ and orientation $\oan$ of the \mbox{$n$-th} anchor, as well as agent-anchor distance $\d = \|\pag-\pan\|$ and direction $\e = (\pag-\pan) / \d$. Thereby, $\oag$, $\oan$, and $\e$ are unit vectors. The anchor topology $\p_1,\o_1,\ldots,\p_N,\o_N$ is assumed to be known accurately. 
Based on this model, the signal term in \eqref{eq:MeasModel}
\begin{align}
\s = \f{\coeff}{d_n^3} \ban\Tr \oag
\label{eq:DipoleModel}
\end{align}
where the inner product $\ban\Tr \oag \in [-1,1]$ describes coil alignment \cite{DumphartPIMRC2016}. We refer to
\begin{align}
\ban &= \Big(\,\f{3}{2} \e\e\Tr - \f{1}{2} \eye_3 \Big) \oan \ , &
\frac{1}{2} &\leq \|\ban\| \leq 1
\end{align}
as the (unitless and virtual) scaled magnetic field at $\pag$ due to the $n$-th anchor. Furthermore, the constant 
\begin{align}
\coeff = \f{\omega\,\mu S\ag S\anc N\ag N\anc}{4\pi\sqrt{R_\mathrm{ag} R_\mathrm{anc}}} \sqrt{P_\mathrm{T}}
\label{eq:coeff}
\end{align}
with permeability $\mu$, coil surface areas $S\ag$ and $S\anc$, coil turn numbers $N\ag$ and $N\anc$, and coil resistances $R_\mathrm{ag}$ and $R_\mathrm{anc}$.
Note that within our model $\coeff^2$ is the received signal power over a coaxially aligned link at $1\,\mathrm{m}$ distance.
A key quantity for link design and localization performance is the signal-to-noise ratio $\mathrm{SNR}_n = \s^2\,/\,\sigma^2$. 

%\vspace{-2mm}

%Alignment factor $\J = \ban\Tr \oag$ is the \mbox{$\oag$-directed} component of the (unitless) scaled magnetic field
%\begin{align}
%\ban &= \Big(\,\f{3}{2} \e\e\Tr - \f{1}{2} \eye_3 \Big) \oan \ , &
%\frac{1}{2} &\leq \|\ban\| \leq 1
%\end{align}
%due to the $n$-th anchor at $\pag$. Consequently, $|\J| \leq \|\ban\|$ and $-1 \leq \J \leq 1$. The magnitude $\|\ban\| = 1$ on the anchor axis and $\|\ban\| = 1/2$ on the plain of the anchor coil. Note that $\ban$ is just a notational model and does not imply that the agent senses fields generated by the anchors (cf. symmetry of mutual inductance). The behavior of $\J$ was discussed in detail in \cite{DumphartPIMRC2016}.

%A key quantity for link design is the signal-to-noise ratio $\mathrm{SNR}_n = \s^2\,/\,\sigma^2$. 
%\begin{align}
%\mathrm{SNR}_n = \f{\s^2}{\sigma^2} = \f{\omega^2 \mu^2 S\ag^2 S\anc^2 N\ag^2 N\anc^2}{16\pi^2 R\ag R\anc} \, \f{P_\mathrm{T}}{N_0 B} \, \f{\J^2}{\d^{6}} \ .
%\end{align}
%Roughly speaking, $\mathrm{SNR}_n$ should be $10\,\mathrm{dB}$ or higher for observation $\y$ to contribute to accurate 3-D localization.% (this was seen in omitted numerical experiments).
%NACH HINTEN ZUM PLOT?

%% file: sections/03-CRLB.tex
The CRLB on the root mean square (RMS) error of a position estimator is a well-established tool for the study of localization performance \cite{ShenTIT2010} and a popular benchmark for practical localization algorithms. For this purpose, this section derives the CRLB for the given near-field position estimation problem.

We want to estimate agent position $\pag$ from the observations $\y$. In doing so we must consider the unknown $\oag$ as nuisance parameter because it affects the statistics of $\y$. Thus, we estimate $\pag$ and $\oag$ jointly. In order to bypass the constraint $\|\oag\| = 1$, we choose the standard spherical parametrization
\begin{align}
\oag = \left[\ \cos\phi\sin\theta\, , \ \sin\phi\sin\theta\, , \ \cos\theta \ \right]\Tr
\label{eq:OSpherical}
\end{align}
with azimuth angle $\phi$ and polar angle $\theta$. Therewith, the estimation parameter of interest is the 5D vector
\begin{align}
\bp := [\,\pag\Tr\,,\,\phi\,,\,\theta\ ]\Tr \ . 
\end{align}

Signal model \eqref{eq:MeasModel} is of the form $\y = \s(\bp) + \w$ where $\s(\bp)$ are deterministic functions of $\bp$, defined by the coupling model \eqref{eq:DipoleModel} and the known anchor topology. This simple signal-in-AWGN structure has the following convenient estimation-theoretic consequences: \cite{Kay1993}
\begin{itemize}
\item The log-likelihood function (without constant term)
\begin{align}
L(\bp) = -\f{1}{2 \sigma^2} \sum_{n=1}^N \big(\s(\bp) - \y \big)^2 .
\label{eq:logLHF}
\end{align}
\item The associated $5\times 5$ Fisher information matrix
\begin{align}
\FIM_\bp = \f{1}{\sigma^2} \sum_{n=1}^N \fp{\s}{\bp} \Big( \fp{\s}{\bp} \Big)\Tr .
\label{eq:FIM}
\end{align}
\end{itemize}

%\textcolor[rgb]{0,0.58,0}{Comment on: \eqref{eq:FIM} is a sum of rank $1$ matrices, so using $N \geq 5$ anchors is a necessary condition for $\FIM_\bp$ to be nonsingular. This is consistent with the 5 equations for 5 unknowns blabla.}

Computing $\FIM_\bp$ for some $\bp$ requires the geometric gradient $\partial\s/\partial\bp$ which is expanded in Appendix \ref{app:gradient}. The CRLB on the variance of an unbiased estimator $\hat\bp$ states $\text{var}\{ \hat \param_i \big\} \!\geq\! (\FIM_{\bp}^{-1})_{i,i}\,$. 
The mean squared error of an unbiased position estimator $\hat\p\ag$ is thus bounded by \cite{ShenTIT2010} 
\begin{align}
% & \text{var}\{ \hat{p}_\mathrm{x} \big\} + \text{var}\{ \hat{p}_\mathrm{y} \big\} + \text{var}\{ \hat{p}_\mathrm{z} \big\} \\ &=
% \text{E}\{ (\hat{p}_\mathrm{x} -  p_\mathrm{x})^2 + (\hat{p}_\mathrm{y} -  p_\mathrm{y})^2 + (\hat{p}_\mathrm{z} -  p_\mathrm{z})^2 \big\} \\ &=
\text{E}\big\{\|\hat\p\ag -  \pag\|^2 \big\} \geq \text{tr}\big\{ \big(\FIM_{\bp}^{-1}\big)_{1:3,1:3} \big\} \, .
\label{eq:SPEB}
\end{align}
%Analogously, the accuracy of unbiased orientation estimators is limited by $\text{var}\{ \hat\phi   \} \geq ( \FIM_{\bp}^{-1} )_{4,4}$ and $\text{var}\{ \hat\theta \} \geq ( \FIM_{\bp}^{-1} )_{5,5}$. KICK OUT?\\
The resulting lower bound on the RMS position error is usually referred to as position error bound (PEB) \cite{ShenTIT2010}. We denote
\begin{align}
\mathrm{PEB}(\,\pag,\oag) = \sqrt{\text{tr}\big\{ \big(\FIM_{\bp}^{-1}\big)_{1:3,1:3} \big\}} \ .
\label{eq:PEB}
\end{align}

%% file: sections/04-Eval.tex
In this section we evaluate the performance limits of near-field localization for realistic technical parameters in a square room of $3\,\mathrm{m}$ height. The anchors are installed to the side walls in the pattern indicated in Figure~\ref{fig:CoilSetup} which aims at a large spread in all three dimensions (we do not address optimal anchor deployment in this paper). All relevant technical parameters are summarized in Table~\ref{tab:params}. In particular, we choose the ISM band at $13.56\,\mathrm{MHz}$ used by popular NFC and RFID standards. $P_\mathrm{t}$ is set to $10\,\mathrm{mW}$, a typical transmit power for mobile consumer devices. The agent coil parameters are according to the data sheet of a market standard NFC antenna \cite{AgentAntenna}. The anchor coils have 50 turns and rectangular shape with rather compact side lengths of $150\,\mathrm{mm}$ and $100\,\mathrm{mm}$. We calculated $R_\mathrm{anc} = 17\,\Omega$ for copper wire of $0.5\,\mathrm{mm}$ thickness, considering skin and proximity effects and radiation resistance. In terms of SNR, the receive filter bandwidth $B$ should be as small as possible. We set $B = 500\,\mathrm{Hz}$ in order to avoid expensive requirements on frequency synchronization and to allow for movement tracking (the update rate is limited by receive filter transients, which die out on the time scale of $1/B = 2\,\mathrm{ms}$).

\begin{table}
\centering
\renewcommand{\arraystretch}{1.2}
\begin{tabular}{c|c|l}
\textbf{Quantity} & \textbf{Value} & \multicolumn{1}{c}{\textbf{Comment}}\\\hline
$\mu$  & $4\pi \cdot 10^{-7}\,\mathrm{H}/\mathrm{m}$ & permeability (vacuum) \\
$\omega$ & $2\pi \cdot 13.56\,\mathrm{MHz}$ & angular frequency (ISM band) \\
$S\ag$   & $50\,\mathrm{mm} \cdot 35\,\mathrm{mm}$ & rectangular surface area \cite{AgentAntenna} \\
$N\ag$   & $4$  & turn number (agent) \cite{AgentAntenna} \\
$R\ag$   & $4\,\Omega$ & coil resistance (agent) \cite{AgentAntenna} \\
$S\anc$  & $150\,\mathrm{mm} \cdot 100\,\mathrm{mm}$ & rectangular surface area \\
$N\anc$  & $50$ & turn number (anchor) \\
$R\anc$  & $17\,\Omega$ & coil resistance (anchor) \\
$P_\mathrm{t}$ & $10\,\mathrm{dBm}$ & transmit power \\
$T$ & $300\,\mathrm{K}$ & room temperature \\
$B$ & $500\,\mathrm{Hz}$ & receive filter bandwidth \\
$F$ & $8\,\mathrm{dB}$ & receiver noise figure \\\hline
\multicolumn{3}{c}{Resulting Powers} \\\hline
$\rho^2$ & $-50.4\,\mathrm{dBm}$ & Rx power: $\d = 1\,\mathrm{m}$, coax. \\ % cf. Section \ref{sec:model} \\
$\sigma^2$ & $-128.8\,\mathrm{dBm}$ & thermal noise floor %\\
\end{tabular}
\vspace{3mm}
\caption{Technical parameters used in all simulations.}
\label{tab:params}
\vspace{-3mm}
\end{table}

\begin{figure}[!ht]
	\vspace{-1.2mm}
  \centering
  \psfrag{sn}{\hspace{-18.3mm}\raisebox{1mm}{\footnotesize Received Signal Power $\s^2$ $[\mathrm{dBm}]$}}
  \psfrag{dn}{\hspace{-12mm}\raisebox{-1mm}{\footnotesize Coil Separation $\d \ [\mathrm{m}]$}}
  %\psfrag{r2}{\hspace{-33.5mm}\raisebox{-.3mm}{\footnotesize\textcolor[rgb]{1,.616,0}{$\coeff^2$}}}
  \psfrag{r2}{\hspace{-31.3mm}\raisebox{-.4mm}{\scriptsize $\coeff^2 \ldots$ Rx power over $1\,\mathrm{m}$ coaxial link}}
  \psfrag{ma}{\hspace{0mm}\raisebox{-.1mm}{\scriptsize $\sigma^2 + 10\,\mathrm{dB}$}}
  \psfrag{s2}{\hspace{5.5mm}\raisebox{-.3mm}{\scriptsize\textcolor[rgb]{1,0,0}{$\sigma^2$}}}
  %\psfrag{s2}{\hspace{4mm}\raisebox{-.3mm}{\footnotesize$\sigma^2$}}
  \psfrag{loc}{\hspace{2.1mm}\raisebox{-.8mm}{\scriptsize robust}}
  \psfrag{fea}{\hspace{2.1mm}\raisebox{-1.2mm}{\scriptsize ranging}}
  \psfrag{com}{\hspace{-.5mm}\raisebox{.2mm}{\scriptsize\textcolor[rgb]{1,0,0}{outage}}}
  \psfrag{out}{\hspace{0mm}\raisebox{-.1mm}{\footnotesize\textcolor[rgb]{1,0,0}{}}}
  \psfrag{mi1}{\hspace{-.4mm}\raisebox{-.6mm}{\scriptsize\textcolor[rgb]{0,0,1}{coils}}}
  \psfrag{mi2}{\hspace{-.4mm}\raisebox{-.9mm}{\scriptsize\textcolor[rgb]{0,0,1}{misaligned}}}
  \psfrag{co1}{\hspace{-4.3mm}\raisebox{-.1mm}{\scriptsize\textcolor[rgb]{0,0,1}{coils}}}
  \psfrag{co2}{\hspace{-4.3mm}\raisebox{-.4mm}{\scriptsize\textcolor[rgb]{0,0,1}{coaxial}}}
  \includegraphics[width=\columnwidth,trim=3 -5 17 10,clip=true]{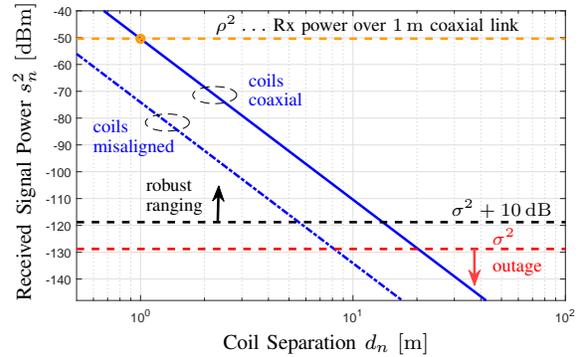} % -2, 18
  \vspace{-5.2mm}
  \caption{The plot shows received signal power over distance. The misaligned case is set to a loss of $-23.7\,\mathrm{dB}$, following from the \mbox{$10\%$-th} percentile coupling value for random coil orientations \cite{DumphartPIMRC2016}.}
	\label{fig:Powers}
\end{figure}

Figure~\ref{fig:Powers} shows a simple evaluation of received power over distance and thereby provides a rough estimate of the usable operation range with the chosen setup parameters. For $\d > 20\,\mathrm{m}$, received signal power $\s^2$ drops below $\sigma^2$ even for a coaxially aligned link (i.e. $\ban\Tr \oag = 1$), thus rendering accurate localization infeasible over such range. However, at distances around $5\,\mathrm{m}$ or less, we experience $\mathrm{SNR}_n > 10\,\mathrm{dB}$ even for poor coil alignment, which is eligible for accurate localization.

\begin{figure}[!ht]
	\centering
  \psfrag{PEB}{\hspace{-8.6mm}\raisebox{1mm}{\footnotesize Median PEB $[\mathrm{m}]$}}
  \psfrag{DXY}{\hspace{-11.5mm}\raisebox{-1.4mm}{\footnotesize Room Side Length $[\mathrm{m}]$}}
  \psfrag{oagax}{\hspace{0mm}\raisebox{0mm}{\scriptsize $\,\oag\!$ arbitrary\! \&\! unknown}}
  \psfrag{oagzxxxxxxxxxxxxxxxxxxxxxxxxxx}{\hspace{0mm}\raisebox{0mm}{\scriptsize $\,\oag\!$ vertical\! \&\! known}}
  \psfrag{nval}{\hspace{-2.5mm}\raisebox{1.0mm}{\scriptsize $N\!=\! 5, 8, 12, 20, 40$}}
  \includegraphics[width=\columnwidth,trim=17 -6 38 0,clip=true]{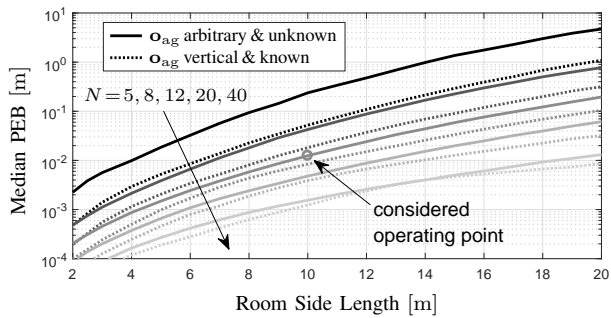}
  \vspace{-5.7mm}
  \caption{Position error bound (PEB) over room side length for different numbers of anchors. The considered room has a square floor plan and $3\,\mathrm{m}$ height. The agent has uniformly distributed position within this room. We compare the 5D case with unknown agent orientation, drawn from a uniform distribution, to the 3D case with known $\oag = [0\ 0\ 1]\Tr$.}
	\label{fig:RoomSizeCRLB}
  %\vspace{-3mm}
\end{figure}

Henceforth, we assume agent position $\pag$ to be random with uniform distribution within the room and, likewise, uniformly distributed agent orientation $\oag$. Consequently, the PEB \eqref{eq:PEB} becomes a random variable. Figure~\ref{fig:RoomSizeCRLB} shows the median PEB versus room side length for different numbers of anchors, whereby each data point was determined empirically by simulation of random agent deployments. As expected, position errors are lowest in small rooms with many anchors. To highlight the effect of an unaligned agent coil, the plot also shows the median PEB for the case of known $\oag$ fixed in vertical direction. Unsurprisingly, this setup performs better, however the difference is minor when a sufficient number of high-SNR observations is available.

\vspace{-1mm}
\begin{figure}[!ht]
	\centering
	\psfrag{ECDF}{\hspace{-5.0mm}\raisebox{0.0mm}{\scriptsize Empirical CDF}}
	\psfrag{RMSPEm}{\hspace{-6.7mm}\raisebox{-0.6mm}{\scriptsize RMS Position Error $[\mathrm{m}]$}}
	\psfrag{RMSAEd}{\hspace{-7.2mm}\raisebox{-0.6mm}{\scriptsize RMS Angle Error $[ \, \degree \, ]$}}
  \includegraphics[width=.95\columnwidth,trim=30 -20 38 22,clip=true]{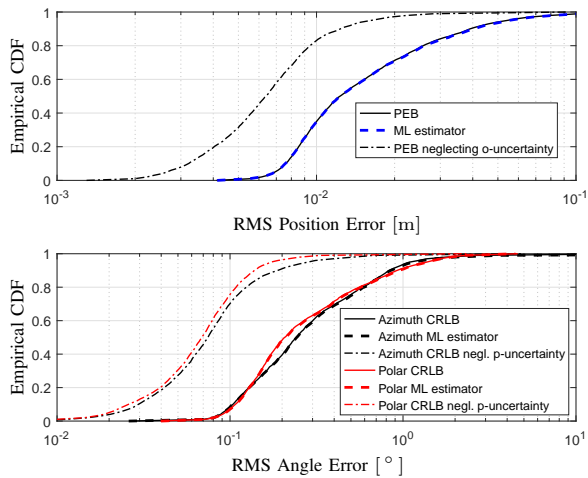}
  \vspace{-6.5mm}
  \caption{The plots show the statistics of CRLB and ML estimator in terms of RMS error of position and orientation angle estimators, respectively, for random agent deployment in a $10\,\mathrm{m} \times 10\,\mathrm{m} \times 3\,\mathrm{m} $ room using $12$ anchors. Also shown are naive bounds which ignore the uncertainty in the respective other domain.}
	\label{fig:CRLB}
  %\vspace{-3.0mm}
\end{figure}

We identify the $N=12$ case with $10\,\mathrm{m}$ side length, which shows a median PEB of $1.2\,\mathrm{cm}$, as an attractive use case with reasonable infrastructure cost. We will use this operating point for the remainder of the paper.
Figure~\ref{fig:CRLB} shows the statistics of the PEB at the chosen operating point. Note that $10\,\mathrm{cm}$ accuracy is feasible for almost any agent deployment. The PEB is compared to the ML estimator, which was computed from 1000 noise realizations per agent deployment, and their statistics match to a high degree. The comparison to a naive PEB, which is obtained \cite{ShenTIT2010} by exchanging $(\FIM_{\bp}^{-1} )_{1:3,1:3}$ with $( (\FIM_{\bp})_{1:3,1:3} )^{-1}$ in \eqref{eq:PEB} and has the effect of assuming $\oag$ perfectly known, highlights the impact of orientation uncertainty: the PEB is approximately twice the naive PEB in the median. The lower plot depicts the corresponding statistics of the orientation angles. Azimuth and polar angle show similar behavior, with an accuracy better than $1^\circ$ for about $91\,\%$ of deployments. Here, the ratio of proper to naive median bound exceeds three. This indicates that orientation estimation is more sensitive to position uncertainty than vice versa.

%\ifstuff
%\begin{figure}[!ht]
	%\centering
	%\includegraphics[width=\columnwidth,trim=40 8 35 24,clip=true]{figures/SNR}
  %\caption{Anchor SNR statistics for the chosen anchor topology in our room of $10\,\mathrm{m} \times 10\,\mathrm{m} \times 3\,\mathrm{m}$, with uniformly random position and orientation within this room. \textcolor[rgb]{0,0.58,0}{Explain that these are key design metrics for the anchors setup and tech parameters. Turn SNR-PDFs into CDFs maybe.} \textcolor[rgb]{1,0,0}{Computed with outdated params.}}
	%\label{fig:SNR}
%\end{figure}
%\fi

%% file: sections/05-Algos.tex
\subsection{Maximum-Likelihood Estimation}
\label{sec:subML}
The ML estimator $\hat\bp_\mathrm{ML} = \argmax_{\bp} L(\bp)$ chooses the parameter value which maximizes log-likelihood function \eqref{eq:logLHF}. Thus, its computation amounts to solving a non-linear least squares problem
\begin{align}
\hat\bp_\mathrm{ML} = \argmin_{\bp} \sum_{n=1}^N \left(\s(\bp) - \y\right)^2 \ .
\label{eq:nonlinlsq}
\end{align}
Due to the lack of a closed-form solution, we attempt to compute $\hat\bp_\mathrm{ML}$ by numerical minimization of the non-convex cost function. In particular, we define an estimation algorithm termed $\text{ML}_{5\mathrm{D}}$ as the application of a trust-region non-linear least squares solver%
\footnote{In particular, we use the Matlab\textsuperscript{TM} function \texttt{lsqnonlin} with the \texttt{trust-region-reflective} option. This requires the error Jacobian in closed form and leads to slight convergence advantages over the \texttt{levenberg-marquardt} option. We will use the latter for $\text{ML}_{3\mathrm{D}}$ and \AlgoShort{} because the associated Jacobians are unavailable.}
to \eqref{eq:nonlinlsq}, using a certain $\bp$ as initialization. An estimate obtained this way can differ severely from the actual $\hat\bp_\mathrm{ML}$ whenever the solver converges to a local minimum \cite{ColemanSIAM1996}. The solver is provided the $5\times N$ Jacobian holding all geometric error gradients $\partial (\s - \y) / \partial\bp = \partial s_n / \partial\bp$, which are given in closed form in Appendix \ref{app:gradient}. 

We can emulate $\hat\bp_\mathrm{ML}$ by initializing $\text{ML}_{5\mathrm{D}}$ at the true $\bp$, which poses a useful performance benchmark.\footnotemark   
\footnotetext{We will not use the PEB directly as a benchmark because it applies to the RMS error, which in turn is an unsuitable measure for estimation algorithms that suffer from ambiguities due to local extrema.}
For this purpose, it is important to clarify whether $\hat\bp_\mathrm{ML}$ attains the CRLB, which is unclear because $\hat\bp_\mathrm{ML}$ may be biased\footnotemark \cite{Kay1993}.
\footnotetext{The bias $\mathrm{E}[\hat\bp_\mathrm{ML}-\bp]$ was found to be several orders of magnitude smaller than noise-induced RMS errors in omitted experiments. It is thus negligible in terms of error performance.}%
Figure~\ref{fig:CRLB} however shows empirically that $\hat\bp_\mathrm{ML}$ does indeed attain the CRLB for our problem. 

A later section will demonstrate that $\text{ML}_{5\mathrm{D}}$ with random initialization shows poor convergence behavior in terms of speed and global optimality and is thus unsuitable for fast and accurate localization.

\subsection{Parameter Space Reduction from 5D to 3D}

In this section, we treat the nuisance parameter $\oag$ separately in a rigorous way, in order to alleviate the problem of high dimensionality in \eqref{eq:nonlinlsq}.
%In this section, we eliminate the orientation nuisance parameter from the search space in a rigorous fashion, in order to alleviate problems caused by the complicated 5D cost function \eqref{eq:nonlinlsq}.

We collect the anchor-agent distances in matrix \mbox{$\D_\p = \diag\{d_1(\p), \ldots ,d_N(\p) \} \in \mathbb{R}^{N\times N}$} and the scaled magnetic fields in $\B_\p = [\banNoIdx_1(\p),\ldots,\banNoIdx_N(\p) ] \in \mathbb{R}^{3\times N}$. They are denoted as functions of an arbitrary position hypothesis $\p$ as they follow from the anchor topology for any $\p$. Yet, $\textbf{y} = [y_1, \ldots, y_N]\Tr$ relates to the true $\pag$ and $\oag$. We express signal model \eqref{eq:MeasModel}, \eqref{eq:DipoleModel} in vector form
\begin{align}
\textbf{y} &= \coeff\,\D_{\pag}^{-3} \B_{\pag}\Tr \oag + \wvec \ .
\label{eq:yVec}
\end{align}
Therewith, $(\hat\p_{\mathrm{ag}} , \hat\o_\mathrm{ag} ) = \argmin_{\p,\o} \|\, \coeff\,\D_\p^{-3} \B_\p\Tr \o - \textbf{y} \|^2$ subject to $\|\o\|^2 = 1$ is the ML estimate, equivalent to the previous section. We note that, given any (temporarily fixed) position hypothesis $\p$, we can compute the ML orientation estimate
\begin{align}
\hat\o_\p = \argmin_{\o} \|\, \coeff\,\D_\p^{-3} \B_\p\Tr \o - \textbf{y} \|^2 \ \text{ s.t. }\ \|\o\|^2 = 1
\label{eq:oStepUnscaled}
\end{align}
which allows us to reformulate the ML estimator as
\begin{align}
\hat\p\ag = \argmin_{\p} \|\, \coeff\,\D_\p^{-3} \B_\p\Tr \hat\o_\p - \textbf{y} \|^2 \ .
\label{eq:pStepUnscaled}
\end{align}
This way, we transformed the 5D minimization problem \eqref{eq:nonlinlsq} into an alternating minimization procedure %\footnote{The scheme is inspired by the expectation–maximization algorithm.} 
consisting of a 3D $\p$-step and a 2D $\o$-step: an iterative solver applied to \eqref{eq:pStepUnscaled} requires the computation of $\hat\o_\p$ through \eqref{eq:oStepUnscaled} after every update of position hypothesis $\p$. 

The $\o$-step \eqref{eq:oStepUnscaled} is a linear least squares problem with quadratic equality constraint. This problem can be solved efficiently as follows, using the theory presented in \cite{Gander1980}. Let
$\mathbf{A} = \rho \D_\p^{-3} \B_\p\Tr \in \mathbb{R}^{N \times 3}$.
By considering the stationary points of the Lagrange function associated with \eqref{eq:oStepUnscaled}, we find the orientation given Lagrange multiplier $\lambda$
\begin{align}
\hat\o_\p(\lambda) = (\mathbf{A}\Tr \mathbf{A} + \lambda\eye_3 )^{-1} \mathbf{A}\Tr \mathbf{y} \ .
\label{eq:oFromLambda}
\end{align}
Let $\lambda^*$ denote the largest $\lambda$ satisfying the constraint $\|\hat\o_\p(\lambda)\|^2 = 1$. Then $\hat\o_\p(\lambda^*)$ is the solution\footnotemark\ to \eqref{eq:oStepUnscaled}.
\footnotetext{We do not elaborate on special cases that occur with probability zero for our noisy estimation problem. For details refer to \cite{Gander1980}.}%
In order to find $\lambda^*$, we use a reformulation \cite{Gander1980}
\begin{align}
\|\hat\o_\p(\lambda)\|^2 = \sum_{i = 1}^3 \f{\mu_i c_i^2}{(\mu_i + \lambda)^2} = 1
\label{eq:lambdaSumFormulation}
\end{align}
based on the eigenvalue decomposition of rank 3 matrix $\mathbf{A}\mathbf{A}\Tr = \sum_{i=1}^3 \mu_i \mathbf{u}_i \mathbf{u}_i\Tr$ and $c_i = \mathbf{u}_i\Tr \mathbf{y}$. Therewith, we can compute $\lambda^*$ efficiently by finding the largest real root of a sixth-order polynomial in $\lambda$ which arises from multiplication of \eqref{eq:lambdaSumFormulation} with its three denominators.

Because of the very efficient and reliable $\o$-step resulting from this method, only the 3D minimization problem \eqref{eq:pStepUnscaled} remains as a computational challenge. Thus, we effectively reduced the initial problem from 5D to 3D. 

We define the $\text{ML}_{3\mathrm{D}}$ algorithm as the application of the Levenberg-Marquardt solver to \eqref{eq:pStepUnscaled}, using a certain initial $\p$. Later in the paper, we will see that $\text{ML}_{3\mathrm{D}}$ has much faster convergence speed than $\text{ML}_{5\mathrm{D}}$, owing to the dimensionality reduction. The robustness improvement over $\text{ML}_{5\mathrm{D}}$ is minor though: $\text{ML}_{3\mathrm{D}}$ also does not converge to the global optimum reliably.

\subsection{\AlgoFull{} (\AlgoShort{}) Algorithm}
The $\text{ML}_{3\mathrm{D}}$ algorithm of the previous subsection still suffers from local cost function extrema. We attribute part of the problem to the high dynamic measurement range due to path loss: the strongest $\y$-values dominate the squared error in \eqref{eq:pStepUnscaled} at most position hypotheses. As a result, the majority of anchors is effectively ignored in early solver iterations, which hinders convergence to the global optimum. Apparently this problem could be avoided with a more balanced cost function.

On that note, we apply a distance-dependent scaling
\begin{align}
\jh = \f{1}{\coeff}\,\D_\p^{3}\,\textbf{y}
\label{eq:scaling}
\end{align}
and, as an extension of \eqref{eq:pStepUnscaled}, we propose position estimation based on the (weighted) cost function
\begin{align}
\hat\p\ag = \argmin_{\p} \big\|\, \B_\p\Tr \hat\o_\p - \jh \big\|^2 \ .
\label{eq:pStepScaled}
\end{align}
In the following, we refer to the application of the Levenberg-Marquardt non-linear least squares solver to \eqref{eq:pStepScaled} as \AlgoFull{} (\AlgoShort{}) algorithm. 
Like in the $\text{ML}_{3\mathrm{D}}$ case, we compute $\hat\o_\p$ for each solver iteration of \eqref{eq:pStepScaled}, with the difference that we apply scaling \eqref{eq:scaling} also to the $\o$-step. In particular, we require
\begin{align}
\hat\o_\p  = \argmin_{\o} \big\|\, \B_\p\Tr \o - \jh \big\|^2 \ \ \text{  s.t.  } \ \ \|\o\|^2 = 1 \ .
\label{eq:oStepScaled}
\end{align}
This is solved analogously to \eqref{eq:oStepUnscaled} of the previous section: compute $\hat\o_\p(\lambda^*)$ with \eqref{eq:oFromLambda} after finding the largest real root $\lambda^*$ of the sixth-order polynomial associated with \eqref{eq:lambdaSumFormulation}, but set $\textbf{A} = \B_\p\Tr$ and $c_i = \mathbf{u}_i\Tr \jh$ instead.%\footnotemark
%\footnotetext{We studied a modification to \AlgoShort{} which ignores the norm constraint while computing $\hat\o_\p = \argmin_{\o} \big\| \B_\p\Tr \o - \jh \big\|^2$ with a QR solver but then employs $\hat\p\ag = \argmin_{\p} \big\|\, \B_\p\Tr \hat\o_\p - \jh \big\|^2 + \big( \|\hat\o_\p\| - 1 \big)^2$ with an additional cost function term that accounts for the norm constraint. For the considered setup, this approach led to the same observed accuracy and robustness as \AlgoShort{}, although with slightly faster execution time.}

The idea of scaling \eqref{eq:scaling} is mapping all observations onto a common value range (the elements of $\jpag$ are coil alignment factors $\ban\Tr\oag \in [-1,1]$ plus noise). This aims at relaxing cost functions \eqref{eq:pStepScaled} and \eqref{eq:oStepScaled} by preventing a degenerated value range of their error terms.

The \AlgoShort{} algorithm yields reduced accuracy as compared to $\text{ML}_{3\mathrm{D}}$ and $\text{ML}_{5\mathrm{D}}$ in case of global optimality. This is because \eqref{eq:scaling} scales the noise variances individually and, in consequence, the least squares estimates \eqref{eq:pStepScaled} and \eqref{eq:oStepScaled} are not estimates in the ML sense. However, this drawback is mitigated by cascading \AlgoShort{} and $\text{ML}_{3\mathrm{D}}$ (initialized at the \AlgoShort{} estimate).

% Put major performance plot here to enforce it being on top of page 5.
\begin{figure}[!ht]
	\centering
	\subfloat[Position error performances with one random initialization.]{% 19
	\psfrag{ECDF}{\hspace{-6.3mm}\raisebox{1mm}{\footnotesize Empirical CDF}}
  \psfrag{PEm}{\hspace{-8.1mm}\raisebox{-1.2mm}{\footnotesize Position Error $[\mathrm{m}]$}}
  \includegraphics[width=.95\columnwidth,trim=28 -14 28 0,clip=true]{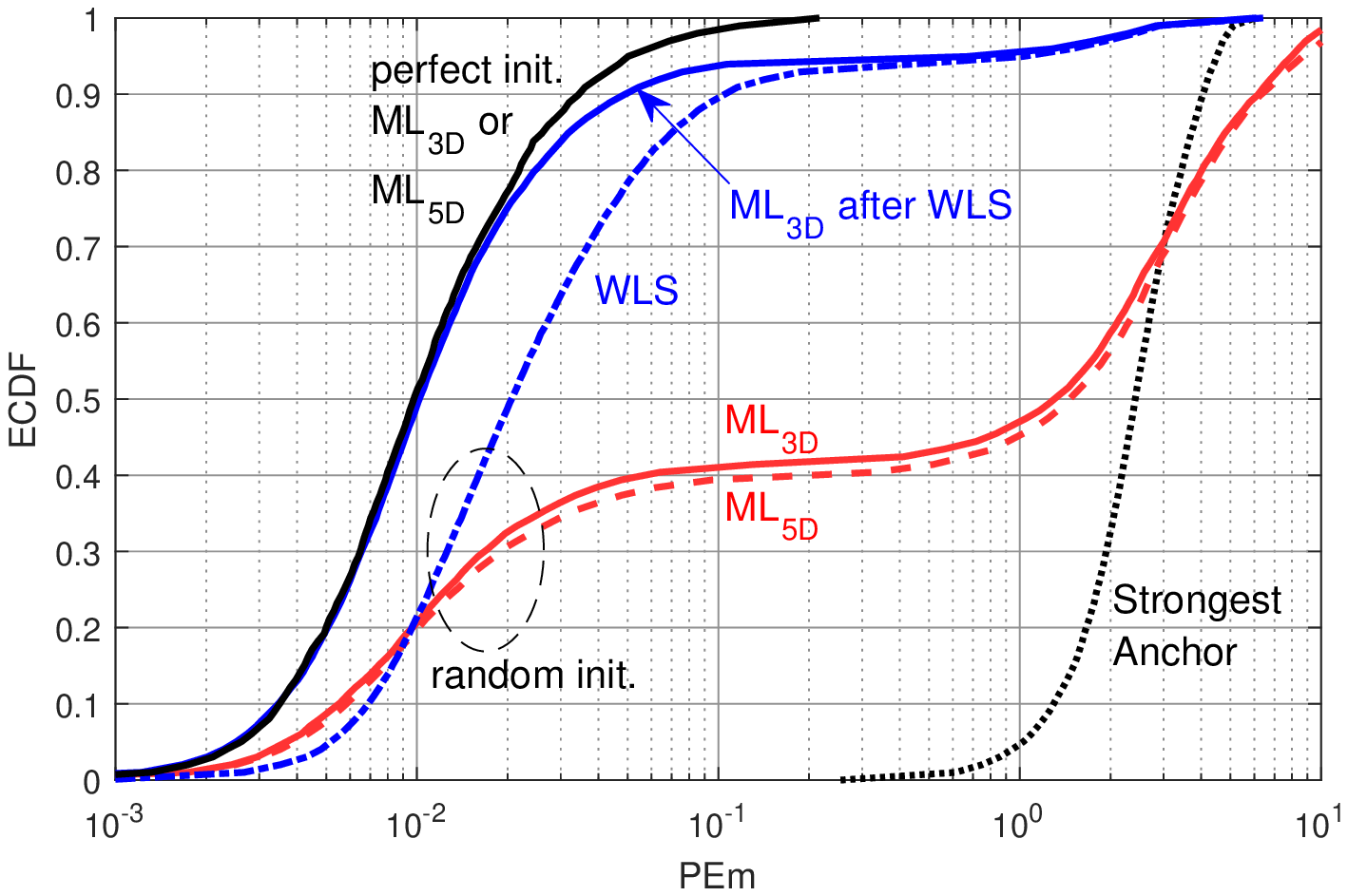}
  \label{fig:PerfCDF1Init}}
  %\vspace{-1mm}\\
  \ \\
  \subfloat[Number of iterations required for solver termination.]{%
  \psfrag{Empirical CDF}{\hspace{-2.6mm}\raisebox{1mm}{\footnotesize Empirical CDF}}
	\psfrag{Solver Iteration Count}{\hspace{-3.1mm}\raisebox{-1.2mm}{\footnotesize Solver Iteration Count}}
	\includegraphics[width=.95\columnwidth,trim=28 -14 28 0,clip=true]{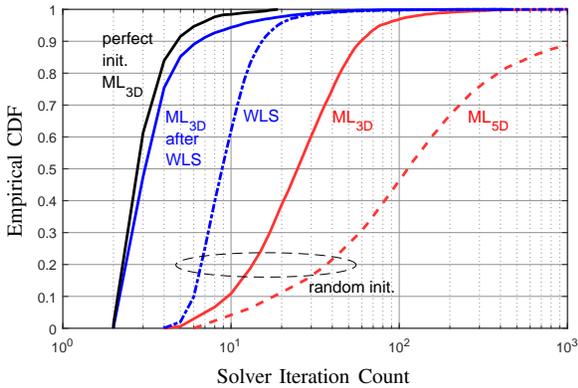}
	\label{fig:Runtime}}
  %\vspace{-1mm}\\
  \ \\
  \subfloat[Position error performances with 3 random initializations.]{%
  \psfrag{ECDF}{\hspace{-6.3mm}\raisebox{1mm}{\footnotesize Empirical CDF}}
  \psfrag{PEm}{\hspace{-8.1mm}\raisebox{-1.2mm}{\footnotesize Position Error $[\mathrm{m}]$}}
  \includegraphics[width=.95\columnwidth,trim=28 -14 28 0,clip=true]{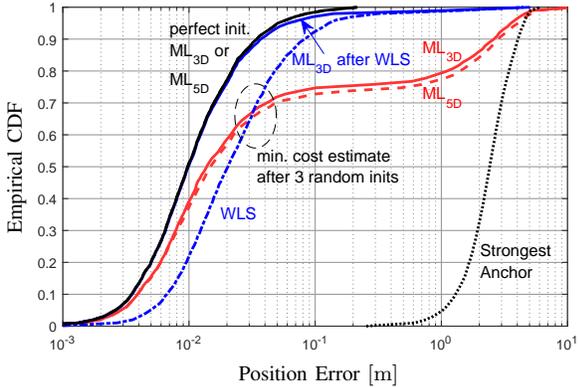}
  \label{fig:PerfCDF3Init}}
  \caption{Numerical comparison of the discussed algorithms for random agent deployment in a $10\,\mathrm{m} \times 10\,\mathrm{m} \times 3\,\mathrm{m}$ room using $12$ anchors.}
	\label{fig:Perf}
  %\vspace{-2mm}
\end{figure}

\subsection{Evaluation of Estimation Performance}
\label{sec:PerfEval}

To complete the picture of the presented algorithms, we compare them numerically for the parameters described in Section \ref{sec:eval}, with the chosen setup of $12$ anchors in a room of $10\times 10\times 3$ meters. We use a maximum of 1000 iterations and a minimum parameter update of $10^{-6}$ as termination criteria for all solvers. Wherever indicated, random initialization refers to uniform sampling of an initial position within the room (and uniformly sampled orientation in the case of $\text{ML}_{5\mathrm{D}}$).\footnotemark 
\footnotetext{We tested several solver initialization heuristics, e.g., choosing the SNR-weighted center of anchor positions. The resulting improvements over random initialization were appreciable but we considered them too insignificant for inclusion, for the sake of clarity.}
The true agent deployment is sampled the same way.

Figures \ref{fig:PerfCDF1Init} and \ref{fig:Runtime} show the error statistics and convergence speed, respectively, of the algorithms with a single initialization. The error statistics are compared to a simple benchmark estimator which just picks the position of the anchor with the strongest received signal, i.e. $\hat\p\ag = \p_{n^*}$ with $n^* = \argmax_n \y^2$.
The randomly initialized $\text{ML}_{5\mathrm{D}}$ is seen to be accurate in only $40\%$ of cases and suffering from local convergence otherwise. Furthermore, it converges slowly with a median iteration count of $113$, and $11\%$ of cases even running $1000$ iterations. $\text{ML}_{3\mathrm{D}}$ improves robustness to a mere $42\%$ but reduces the required iterations to $25$ in the median and to below $100$ in $97\%$ of cases. In comparison, the \AlgoShort{} algorithm exhibits good accuracy with a robustness beyond $94\%$ and fast convergence speed with a median iteration count of just $9$, while $98\%$ of cases require less than $27$ iterations.%
\footnote{With random initialization, we measured mean execution times of $13.7\,\mathrm{ms}$ for \AlgoShort{},  $22.4\,\mathrm{ms}$ for the cascade of \AlgoShort{} and $\text{ML}_{3\mathrm{D}}$, $39.8\,\mathrm{ms}$ for $\text{ML}_{3\mathrm{D}}$, and $331\,\mathrm{ms}$ for $\text{ML}_{5\mathrm{D}}$ for our implementation (no parallelization) running on an Intel Core i7-4790 processor.} 
By cascading \AlgoShort{} and $\text{ML}_{3\mathrm{D}}$ we achieve accuracy at the theoretical limit, constituted by the CRLB-achieving ML estimator, with the $94\%$ robustness of \AlgoShort{}. Similar to perfect initialization, $\text{ML}_{3\mathrm{D}}$ requires just a few iterations after \AlgoShort{}. This indicates that minima of the scaled cost function \eqref{eq:pStepScaled} are close to minima of the original cost function \eqref{eq:pStepUnscaled}.

The remaining non-convexity issues can be addressed by running \AlgoShort{} for several initializations and picking the estimate with the smallest residual cost. Figure \ref{fig:PerfCDF3Init} shows the error statistics for 3 random initializations. We observe a robustness improvement to $98\%$ using \AlgoShort{}.

\ifstuff
\textcolor[rgb]{0,0.58,0}{Comment on lower and upper bounds for $x,y,z$ and why it's smart not to constrain $\phi,\theta$ (could cause problems with local minima on the boundaries, and we do not care about modulo-$2\pi$ ambiguities).}
\fi

%% file: sections/99-Concl.tex
We studied performance regimes and estimation algorithms for near-field 3D localization on the indoor scale with flat coils and arbitrary agent orientation. After deriving the CRLB on the position error and studying its dependence on room size and anchor count, we found $\mathrm{cm}$-accuracy being feasible in a square room of $10\,\mathrm{m}$ side length. To enable such positioning accuracy in practice, we proposed an algorithm which employs a suitable scaling and alternating estimation of position and orientation, with an efficient solution for the orientation step. The resulting localization scheme performs near the CRLB with high robustness and consistently low computational cost. The proposed algorithm is thus a potential enabler for accurate indoor 3D localization with unobtrusive infrastructure and high update rate.

%% file: sections/A1-Circuit.tex
\begin{figure}[!ht]
\centering
\psfrag{M}{\hspace{-1.7mm}\textcolor[rgb]{0,.6,0}{$M_n$}}
\psfrag{ir}{\hspace{-2mm}\raisebox{1.9mm}{\textcolor[rgb]{1,.616,0}{$i_\mathrm{ref}$}}}
\psfrag{Pt}{}
%\psfrag{Pt}{\hspace{-8mm}\raisebox{1.5mm}{$P_\mathrm{tx} = R_\mathrm{tx} i_\mathrm{ref}^2$}}
\psfrag{Rt}{\hspace{-0.35mm}\raisebox{0.5mm}{$R_\mathrm{ag}$}}
\psfrag{Rr}{\hspace{-4mm}\raisebox{0.5mm}{$R_\mathrm{anc}$}}
\psfrag{Lr}{\hspace{-1mm}\raisebox{0.5mm}{$L_\mathrm{anc}$}}
%\psfrag{ZL1}{\hspace{1.3mm}\raisebox{-1.5mm}{$Z_\mathrm{coil}^*$}}
%\psfrag{ZL2}{}
%\psfrag{ZL1}{\hspace{-0.4mm}\raisebox{-1.5mm}{$Z_\mathrm{load}$}}
%\psfrag{ZL2}{\hspace{-1.4mm}\raisebox{-2.7mm}{$=\!Z_\mathrm{coil}^*$}}
\psfrag{ZL1}{\hspace{-2.2mm}\raisebox{.6mm}{\footnotesize Matched}}
\psfrag{ZL2}{\hspace{0.1mm}\raisebox{-.2mm}{\footnotesize LNA}}
\psfrag{vi}{\hspace{-.55mm}\raisebox{-1.2mm}{\textcolor[rgb]{0,0,1}{$v_n^\mathrm{ind}$}}}
\psfrag{vn}{\hspace{-.20mm}\raisebox{-1.2mm}{\textcolor[rgb]{1,0,0}{$v_n^\mathrm{th}$}}}
\psfrag{vL}{\hspace{0.5mm}\raisebox{0.1mm}{$Z_\mathrm{anc}^*$}}
%\psfrag{mk}{\hspace{0mm}\raisebox{0mm}{$R_\mathrm{anc} - j\omega L_\mathrm{anc}$}}
%\psfrag{vL}{\hspace{0.3mm}\raisebox{0.1mm}{\textcolor[rgb]{1,0,1}{$v_n^\mathrm{load}$}}}
%\psfrag{pk}{\hspace{0mm}\raisebox{ .3mm}{\textcolor[rgb]{1,0,1}{$+$}}}
%\psfrag{mk}{\hspace{0mm}\raisebox{-.3mm}{\textcolor[rgb]{1,0,1}{$-$}}}
\psfrag{pk}{}
\psfrag{mk}{}
\psfrag{pb}{\hspace{-.59mm}\raisebox{-0.6mm}{\textcolor[rgb]{0,0,1}{$+$}}}
\psfrag{mb}{\hspace{1.2mm}\raisebox{-0.6mm}{\textcolor[rgb]{0,0,1}{$-$}}}
\psfrag{pr}{\hspace{-.65mm}\raisebox{-0.6mm}{\textcolor[rgb]{1,0,0}{$+$}}}
\psfrag{mr}{\hspace{ .5mm}\raisebox{-0.6mm}{\textcolor[rgb]{1,0,0}{$-$}}}
\includegraphics[width=\columnwidth,trim=0 8 0 0]{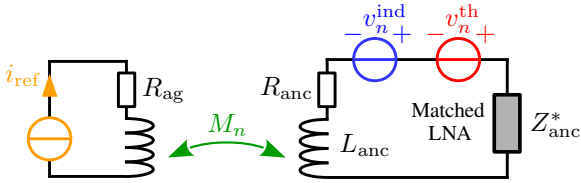}
%\vspace{-5mm}
\caption{Circuit Model of the loosely coupled magneto-inductive link from the agent to the $n$-th anchor.}
\label{fig:CircuitModel}
\end{figure}

This appendix provides a basis of signal model \eqref{eq:MeasModel}, \eqref{eq:DipoleModel}, \eqref{eq:coeff} by means of the circuit description of the magneto-inductive link from the agent to the \mbox{$n$-th} anchor shown in Figure~\ref{fig:CircuitModel}. For the definition of reoccurring quantities we refer to the front matter. 
We employ a loose-coupling assumption which asserts that %antenna impedance shifts due to coupling are negligible. 
antenna impedance do not change appreciably. 
The agent uses active power $P_\mathrm{t} = R_\mathrm{ag} i_\mathrm{ref}^2$ to generate a field which induces a voltage $v_n^\mathrm{ind} = \omega M_n i_\mathrm{ref}$ (effective value) at the anchor coil. The anchor coil is terminated with a low-noise amplifier (LNA) matched to the coil, i.e. its input impedance $Z_\mathrm{LNA} = Z_\mathrm{anc}^* = R_\mathrm{anc} - j\omega L_\mathrm{anc}$. %it shows resistance $R_\mathrm{anc}$ and capacitive reactance for resonance with the inductive coil. 
The signal portion of the power wave into the LNA $s_n = v_n^\mathrm{ind} / \sqrt{4 R_\mathrm{anc}} = \omega M_n  \sqrt{P_\mathrm{t}/ (4 R_\mathrm{ag} R_\mathrm{anc})}\,$.
The mutual inductance $M_n = \frac{\mu_0}{2\pi} S\ag S\anc N\ag N\anc\, \d^{-3} \ban\Tr\oag$ by the dipole model \cite{DumphartPIMRC2016}, which yields \eqref{eq:DipoleModel} and \eqref{eq:coeff}.

The thermal noise voltage $v_{n}^\mathrm{th}$ due to $R_\mathrm{anc}$ and LNA has variance $\mathrm{E}[ (v_{n}^\mathrm{th})^2 ] = 4 N_0 B R_\mathrm{anc}$. The power wave $\w = v_n^\mathrm{th} / \sqrt{4 R_\mathrm{anc}}$ thus exhibits $\mathrm{E}[ \w^2 ] = N_0 B$.

%% file: sections/A2-Gradient.tex
This appendix gives the five-dimensional geometric gradients of the signals $\s$, which are according to the dipole model \eqref{eq:DipoleModel}, at anchors $n = 1,\ldots,N$. In particular, 
\begin{align}
& \fp{\s}{\bp} = \bigg[ \Big( \fp{\s}{\p} \Big)\Tr \, , \, \fp{\s}{\phi} \, , \, \fp{\s}{\theta} \bigg]\Tr
\label{eq:paramGradient}
\end{align}
at some trial position $\p$ and trial orientation $\o$ of spherical angles $\phi$ and $\theta$. It follows from the dipole model \eqref{eq:DipoleModel}. The gradients yield Fisher information matrix \eqref{eq:FIM} and are employed in the $\text{ML}_{5\mathrm{D}}$ algorithm of Section \ref{sec:subML}. The spatial gradient
\begin{align}
& \fp{\s}{\p} = \f{\coeff}{\d^3}\bigg(\fp{\ban\Tr}{\p} - \f{3}{\d}\e\ban\Tr\bigg)\o \ ,
\label{eq:SpatialGradient} \\
& \fp{\ban\Tr}{\p} = \f{3}{2} \f{1}{\d} \Big(\oan\e\Tr + (\oan\Tr\e)(\eye_3 - 2\e\e\Tr)\Big)
%\label{eq:betaJacobian} \ .
\end{align}
follows from the geometric relations $\partial\d/\partial\p = \e$ and $\partial\e\Tr/\partial\p = (\eye_3 - \e\e\Tr) / \d$. On the other hand,
\begin{align}
\fp{\s}{\phi}   &= \f{\coeff}{\d^3} \ban\Tr \fp{\o}{\phi} \ , &
\fp{\s}{\theta} &= \f{\coeff}{\d^3} \ban\Tr \fp{\o}{\theta}
\label{eq:OrientGradient}
\end{align}
%where the derivatives of $\o$-parametrization \eqref{eq:OSpherical} are trivial.
with the derivatives $\partial\o/\partial\phi = [\,-\sin\phi, \, \cos\phi, \, 0 \,]\Tr \sin\theta$ and $\partial\o/\partial\theta = [\, \cos\phi\cos\theta, \, \sin\phi\cos\theta, \, -\sin\theta \,]\Tr$ %, which follow from differentiation
of \eqref{eq:OSpherical}.